\documentclass[reprint,amsmath,amssymb,aps]{revtex4-2}
\usepackage{graphicx}
\usepackage{dcolumn}
\usepackage{bm}
\usepackage{hyperref}
\hypersetup{colorlinks,allcolors=black}

\usepackage{tikz}

\usepackage{wrapfig, blindtext}
\usepackage{arydshln}
\usepackage{pgfplots}
\usepackage{axodraw2}

\pgfplotsset{compat=1.17}
\usepackage{systeme}
\usepackage{tikz}
\usetikzlibrary{positioning,arrows,arrows.meta,shapes}
\usepackage{tcolorbox}
\usepackage{xcolor}
\usepackage{listings}
\usepackage{tikz-3dplot}
\usetikzlibrary{decorations.markings}
\usetikzlibrary{intersections}
\usepackage{capt-of}

\usetikzlibrary{spath3, intersections, hobby}

\begin{document}

\preprint{APS/123-QED}

\title{Vortices, topology and time}

\author{Altay Etkin$^{1,2}$}
\email{altay.etkin22@imperial.ac.uk}
\author{Jo\~ao Magueijo$^1$}%
\email{j.magueijo@imperial.ac.uk}
\author{Farbod-Sayyed Rassouli$^{1,3}$}%
\email{sayyed.rassouli21@imperial.ac.uk}

\affiliation{$^{1}$Theoretical Physics Group, The Blackett Laboratory, Imperial College,
Prince Consort Rd., London, SW7 2BZ, United Kingdom}%
\affiliation{$^{2}$School of Mathematical Sciences and STAG Research Centre, University of Southampton,\\Highfield Campus, Southampton, SO17 1BJ, United Kingdom}
\affiliation{$^{3}$School of Physics and Astronomy, University of Nottingham,\\University Park, Nottingham, NG7 2RD, United Kingdom}

\date{21 June, 2024}

\begin{abstract}
We relate physical time with the topology of magnetic field vortices. We base ourselves on a formulation of unimodular gravity where the cosmological constant $\Lambda$ appears as the canonical dual to a variable which on-shell becomes four-volume time. If the theory is restricted to a topological axionic form (viz. a parity-odd product of an electric and a magnetic field),  such a time variable becomes the spatial integral of the Chern-Simons density. The latter equates to helicity, so that unimodular time is transmuted into the linking number of the vortices of the topological magnetic field, times their flux. With the added {\it postulate}  that this flux is a universal constant, the flow of time can thus be interpreted as the progressive weaving of further links between magnetic field vortices, each link providing a quantum of time with value related to the fixed flux. Non-abelian extensions, and targetting parameters other than $\Lambda$ are briefly examined, exposing different types of vortices and a possible role for inter-linking leading to new phenomenology.
\end{abstract}

\maketitle

\section{Introduction}

The unimodular theory of gravity originates from an issue first raised by Einstein \cite{Einstein-unimod}: should the coordinate invariance of a theory be restricted to volume-preserving coordinate transformations? It was later realized that unimodular gravity is essentially equivalent to General Relativity, with the difference that it demotes the cosmological constant from a pre-given fixed parameter to a constant of motion \cite{unimod1,Kuchar:1991xd,unimod,UnimodLee1,alan,daughton,sorkin1,sorkin2}. Whether this resolves, merely softens, or is irrelevant to the cosmological constant problem remains a disputed matter \cite{weinberg,padilla-review}, outside the scope of this paper. Instead, here we aim to uncover an intriguing connection that has the potential to fundamentally reshape our understanding of time.


As Henneaux and Teitelboim (HT) showed in \cite{unimod}, unimodular gravity is equivalent to a theory with full diffeomorphism invariance, where one adds to the action a term containing a Lagrange multiplier density ${\cal T}^\mu$ enforcing the on-shell constancy of $\Lambda$. This density can be used to build a gauge-invariant physical definition of time, which on-shell becomes ``four-volume time'' \cite{Bombelli,sorkin1,sorkin2,UnimodLee2}. It was noted in \cite{vikman-axion} that the term HT added to the action, by virtue of being something that would be a boundary term were $\Lambda$ to be pre-fixed, can be restricted to an axion-like form. This makes unimodular time take the form of the Chern-Simons (CS) functional of the underlying gauge theory. But the CS functional is nothing but the integral of the helicity density of this underlying topological version of electromagnetism. Hence, it measures the linking number of the magnetic vortices \cite{moffatt_1969,berger1984topological,berger1999introduction}. Time may thus be identified with a topological feature in quantum field theory, and so be discretized. 

In this picture, the cosmological constant is coupled to a topological cousin of electromagnetism, based on an action containing only the parity-odd term $\mathbf{E}\cdot \mathbf{B}$. The fact that time is the canonical dual to $\Lambda$, then relates time to these fields. With the assumption that the magnetic field vortices carry constant flux, time is equated to how the vortices link with each other. The flow of time is therefore the process of weaving in new links between the magnetic vortices and changing their topology. Since the linking number is discrete, each link is a quantum of time.

We exemplify these points with unimodular theory and abelian topological field theory, but the procedure can be generalized to non-abelian theories, with interesting new features (the vortices appear ``colored'').  The HT procedure can also be used as a blueprint for converting into constants of motion parameters other than $\Lambda$ (such as the Planck mass), with alternative conjugate time variables (such as the Ricci time)~\cite{JoaoLetter, JoaoPaper}. In theories where different types of vortices exist (whether due to more generators or more target constants) and their interlinking number becomes relevant, one may expect interesting phenomenology.

\section{Deconstantizing nature's constants}\label{sec1}

``Deconstantization'' is a term introduced in \cite{JoaoPaper, JoaoLetter}, to generalize the HT-unimodular framework pioneered in the seminal work \cite{unimod}, whereby 
constants of Nature that usually appear as fixed parameters in the action are converted into integration constants. Thus, such parameters become constant on-shell only. In this Section, we review this procedure.
The prototype is the HT gravity action~\cite{unimod}:
\begin{equation}\label{fullaction}
    S = S_0+S_U=\frac{1}{2} \int d^4x \ \sqrt{-g} \left(R - 2 \Lambda\right) + \int d^4x \ \Lambda \partial_{\mu} \mathcal{T}^{\mu}.
\end{equation}
The first term is the standard Einstein-Hilbert action with the cosmological constant promoted to a variable to begin with; the second term is the unimodular term, including a Lagrange multiplier, $\mathcal{T}^{\alpha}$, intended to fix $\Lambda$ as a constant, as an equation of motion, and to produce a foliation-dependent time variable. The metric is not employed in the new term, so as not to spoil the Einstein equations; hence the integration uses the coordinate 4-volume element $d^4x$ (without the usual $\sqrt{-g}$ factor) which is a scalar density of weight $-1$. The Lagrange multiplier ${\cal T}^\mu$ is a 4-vector density of weight 1, so that $\nabla_\mu{\cal T}^\mu=\partial_\mu {\cal T}^\mu$ (no connection needed in the covariant derivative), and this divergence is a scalar density of weight 1.  
The action is therefore invariant under the full diffeomorphism group of spacetime coordinate transformations (so that the term ``unimodular'' becomes a misnomer). 
The action is also invariant under the gauge transformation
\begin{equation}\label{gaugetransformation}
    \mathcal{T}^{\mu} \longmapsto \mathcal{T}^{\mu} + \epsilon^{\mu},
\end{equation}
subject to $\partial_{\mu} \epsilon^{\mu}=0$ (so that the gauge transformations have three degrees of freedom). Thus, $\Lambda$ is promoted to a phase space variable, canonically conjugate to the gauge-invariant zero-mode of $\mathcal{T}^0$, known as the unimodular time:
\begin{equation}\label{unimodulartime}
    T_{\Lambda}(t) := \int_{\Sigma_t} d^3x \ \mathcal{T}^0.
\end{equation}
The equations of motions are derived by varying (\ref{fullaction}) with respect to $g_{\mu \nu}$, $\Lambda$ and $\mathcal{T}^{\mu}$:
\begin{align}
    \frac{\delta S}{\delta g^{\mu \nu}} = 0 \ \ \ \ &\Longleftrightarrow \ \ \ \ R_{\mu \nu} = \frac{1}{2} g_{\mu \nu} \bigg(R- 2 \Lambda(x)\bigg),\label{EFE}\\
    \frac{\delta S}{\delta \mathcal{T}^{\mu}} = 0 \ \ \ \ &\Longleftrightarrow \ \ \ \ \partial_{\mu} \Lambda = 0,
    \label{constantalpha}\\
    \frac{\delta S}{\delta \Lambda} = 0 \ \ \ \ &\Longleftrightarrow \ \ \ \ \sqrt{-g} = \partial_{\mu} \mathcal{T}^{\mu}.\label{unimodularequation}
\end{align}
The on-shell dynamics are thus governed by general relativity with $\Lambda$ behaving as the standard cosmological constant. Eq.~(\ref{unimodularequation}) gives us an on-shell expression for $T_{\Lambda}$, identifying it with spacetime volume to the past of $\Sigma_t$, down to a conventional time-zero $\Sigma_0$ hypersurface~\cite{unimod,UnimodLee1,Bombelli}:
\begin{align}\label{deltatime}
    T_{\Lambda}(\Sigma_t) = \int_{\Sigma_0}^{\Sigma_t} d^4x  \sqrt{-g} \equiv V_4(\Sigma_t).
\end{align}
Simple counting of constraints shows \cite{unimod,Isichei:2023epu} that the theory has no new local degrees of freedom. Usually we want time to be intensive, i.e. not to be proportional to the spatial volume, so we replace the above definition by the spatial density:
\begin{equation}\label{intensivetimes}
    T_{\Lambda}(t) = \frac{1}{V_c} \int_{\Sigma_t} d^3x \ \mathcal{T}^0_{\Lambda},
\end{equation}
where $V_c=\int_{\Sigma_t} d^3x$ is the spatial coordinate 3-volume of leaves $\Sigma_t$ (a constant scalar density on the leaves). In the case of unimodular time, $T_\Lambda$, this is thus four-volume per unit of coordinate 
three-volume.

This deconstantization procedure can be generalized to a set of parameters $\boldsymbol{\alpha}$, introducing a corresponding set of densities $\mathcal{T}^{\mu}_{\boldsymbol{\alpha}}$, and an action mimicking the HT-prescription:
\begin{equation}\label{alphaaction}
    S = S_0 + \int d^4x \ \boldsymbol{\alpha} \cdot \partial_{\mu} \mathcal{T}^{\mu}_{\boldsymbol{\alpha}}.
\end{equation}
Thus we define canonical phase space pairs $(\boldsymbol{\alpha},T_{\boldsymbol{\alpha}})$ with the times $T_{\boldsymbol{\alpha}}$  given by
\begin{equation}
    T_{\boldsymbol{\alpha}}(t) := \int_{\Sigma_t} d^3x \ \mathcal{T}^0_{\boldsymbol{\alpha}}.
\end{equation}
In this paper, we will consider the basis:
\begin{equation}\label{alphadef}
    \boldsymbol{\alpha} = (\alpha_1, \alpha_2) = \left(\frac{\Lambda}{8 \pi G},\frac{1}{8 \pi G}\right),
\end{equation}
where $\alpha_1$ is the $\Lambda$-vacuum energy density $\rho_{\Lambda}$, and $\alpha_2$ is the reduced Planck mass squared (or Planck energy) $M^2_{\text{P}}$.
Beside $T_1 \equiv T_\Lambda$ defined above, we have the Ricci time $T_2 \equiv T_R$ which on-shell reads
\begin{equation}\label{TRicci}
    T_2(\Sigma_t) = - \frac{1}{2}
    \int_{\Sigma_0}^{\Sigma_t} d^4x \ \sqrt{-g} R \equiv V_R(\Sigma_t),
\end{equation}
i.e. the Ricci weighted four-volume (see \cite{pad,lombriser2019cosmological} in addition to \cite{Kuchar:1991xd,UnimodLee1,alan,daughton,sorkin1,sorkin2,Bombelli,UnimodLee1}). $T_\Lambda$ and $T_R$ are nothing but the ``fluxes'' used in the sequester model \cite{JoaoPaper,JoaoLetter,Magueijo:2023poa,pad1}.

\section{Abelian axion reduction}\label{sec2}

We consider now the interesting \textit{reduction} of HT gravity pioneered in \cite{vikman-axion} (see also~\cite{Jirousek:2018ago,Jirousek:2020vhy}), with $\alpha_2 = 1$ and variable parameter $\Lambda$. It amounts to replacing the total divergence appearing in the HT-term (cf. (\ref{fullaction})) with the the total divergence appearing in the $\theta$-term
found in axion models. We first consider $U(1)$ as the internal symmetry group (with gauge field $A_{\mu}$ and field strength $F_{\mu\nu}= 2 \partial_{[\mu} A_{\nu]}$), but the extension to non-abelian gauge groups will be discussed later. We thus identify: 
\begin{equation}
    \int d^4x \ \Lambda \partial_{\mu} \mathcal{T}^{\mu}_{\Lambda} = \int d^4x \ \frac{\Lambda}{2} \mathcal{F}^{\alpha \beta} F_{\alpha \beta},
\end{equation}
where $\mathcal{F^{\alpha \beta}} := \frac{1}{2} e^{\mu \nu \alpha \beta} F_{\mu \nu}$ is the densitized dual to $F_{\mu \nu}$ (with density $e_{\alpha \beta \mu \nu}$ defined from $e_{0123}=1$), and since the topological term $\mathcal{F}^{\alpha \beta} F_{\alpha \beta}$ can be written as the divergence of the Chern-Simons current (the dual of the Chern-Simons three-form), we have the identification:
\begin{equation}\label{chernsimonscurrent}
    \mathcal{T}^{\alpha}_{\Lambda} = A_{\beta} \mathcal{F}^{\alpha \beta} = 2 e^{\alpha \beta \mu \nu} A_{\beta} \partial_{\mu} A_{\nu}.
\end{equation}

The $U(1)$ theory is a reduction of HT gravity in that its gauge symmetries ($ A_{\mu} \rightarrow A_{\mu} + \partial_{\mu} \chi$) have a single degree of freedom instead of the three, evident from (\ref{gaugetransformation}). Specifically, we restrict: 
\begin{equation}
    \epsilon^{\alpha}_{\Lambda} =  \partial_{\beta} \left(\chi \mathcal{F}^{\alpha \beta}\right).
\end{equation}
Using the parametrization suggested in \cite{unimod}, where one introduces a general three-vector density $\mathcal{\alpha}^i$ and sets $\epsilon^0_{\Lambda}= - \partial_i \alpha^i$ and $\epsilon^i_{\Lambda} = \dot{\alpha}^i$, this would be equivalent to having $\alpha^i = \chi \mathcal{B}^i$, where  $\mathcal{F}^{0i} = - \mathcal{B}^i := - \mathbf{e}^{ijk} \partial_j A_k$ is the densitized magnetic field and $\mathbf{e}^{123}=1$. These points were made with $\Lambda$ but apply to any other target constant. For example, for the  basis given by (\ref{alphadef}), the symmetry group of the reduced action (\ref{fullaction}) becomes $U(1) \times U(1)$.

The field equation (\ref{unimodularequation}) therefore morphs into:
\begin{equation}\label{unimodularcondition}
    \sqrt{-g} = \frac{1}{2} \mathcal{F}^{\mu \nu} F_{\mu \nu} \equiv 2 E_i \mathcal{B}^i.
\end{equation}
Also, (\ref{constantalpha}) gets replaced with:
\begin{equation}\label{constantlambda}
    \mathcal{F}^{\mu \nu} \partial_{\mu} \Lambda = 0,
\end{equation}
entailing exceptions to $\partial_\mu\Lambda=0$
(a point we return to in Section~\ref{sec5}). Finally, as with any other gauge theory, we have Bianchi identities:
\begin{equation}
    \partial_\mu{\mathcal{F}}^{\mu\nu}=0
\end{equation}
as self-consistency conditions. Unlike with standard electromagnetism (EM), there are no non-homogeneous Maxwell equations, so the analogy with standard EM stops there. This will be important later.

\section{Unimodular time and Vortices}\label{sec3}

We now come to the core remark of this paper. Since (\ref{chernsimonscurrent}) implies $\mathcal{T}^0_{\Lambda} = -A_i \mathcal{B}^i$, we find that, within the axion reduced HT gravity, the unimodular time expressed in (\ref{unimodulartime}) takes the particular expression:
\begin{equation}\label{unimodulartimemagnetichelicity}
    T_{\Lambda}(t) = 
\int_{\Sigma_t} d^3x \ \mathcal{T}^0_{\Lambda} =    
    - \int_{\Sigma_t} d^3x \ A_i \mathcal{B}^i.
\end{equation}
This is nothing but (minus) the {\it helicity} found in fluid dynamics \cite{moffatt_1969}:
\begin{align}
    H &:=  \int_{V} d^3x\, u_i\omega^i = \int_{V} d^3x\ \mathbf{e}^{ijk} u_i \partial_j u_k, 
\end{align}
(where $V$ is a comoving spatial volume) with the potential $A_i$ replaced with the velocity field $u_i$ and the magnetic field $\mathcal{B}^i$ replaced with the vorticity field $\omega^i$. As Moffatt proved in his classic paper \cite{moffatt_1969}, the helicity measures the linking number of {\it vortices}, that is loops of $\omega^i$-lines. Perfect analogues exist in EM and MHD \cite{berger1984topological,berger1999introduction,moffatt1992helicity,arnold2008topological, ricca1998applications}, where vortices refer to loops formed by magnetic flux tubes, and it is in terms of these EM quantities that our helicity may be interpreted, modulo the caveats mentioned at the end of Section~\ref{sec2}.

\begin{figure}[h!]
    \centering

    \includegraphics[width=7cm]{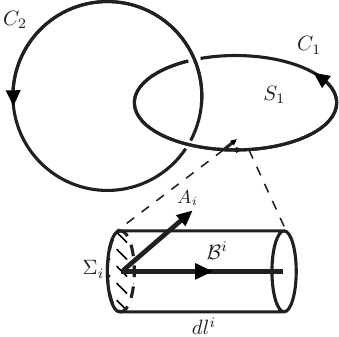}

    \caption{The degree of linkage of two unknotted vortex filaments $C_1$ and $C_2$ illustrated for the case of $\mathcal{B}^i$-lines (but an equivalent argument applies to $\omega^i$-lines). The zoomed inset shows how the helicity volume integral is actually the flux $\Phi= \int d\Sigma_i \mathcal{B}^i$ times the circulation of the vector potential $A_i$. }
    \label{fig:moffatloops}
\end{figure}

The relation between helicity and vortex linking number is well illustrated in the opening example of \cite{moffatt_1969} (which we illustrate in  Fig.\ref{fig:moffatloops} for clarity). Let us start with a single unknotted closed filament $C_1$ of vorticity $\omega^i$ or of magnetic field $\mathcal{B}^i$. The helicity volume-integral over the region $V$ occupied by the vortex is then the constant flux $\Phi$ (of $\omega^i$ or of $\mathcal{B}^i$) along the filament multiplied by the line-integral (circulation) $I$ of the velocity field $u_i$ or of the vector potential $A_i$ along the filament:
\begin{equation}
    \int_{V_1} d^3x \ A_i \mathcal{B}^i= \Phi \int_{C_1} dl^i \ A_i \equiv \Phi I_1.
\end{equation}
This is visually shown in the zoomed inset in Fig.\ref{fig:moffatloops}). 
The latter is nothing but the ``holonomy'' $I_1$, to use the language of gauge theory.  Applying Stokes's theorem to this line-integral, we get the magnetic flux due to other filaments going through any surface bounded by the original filament. Such filaments only intersect this surface if they are topologically linked to the original one. In Fig.\ref{fig:moffatloops} we illustrate this with filament $C_2$ going around $C_1$ once. 
If the other filaments go around $i\in  \mathbb{Z}$ times, their flux is counted $i$-times. Extending this result for $n$-unknotted magnetic vortex lines $C_1,...,C_n$ with fluxes $\Phi_1,...,\Phi_n$, and Gauss linking number $\text{Lk}(C_I,C_J) \in \mathbb{Z}$, we thus have:
\begin{equation}\label{magnetichelicity}
 H = \sum_{I=1}^n \Phi_I I_I = \sum_{I,J=1}^n \text{Lk}(C_I,C_J) \Phi_I \Phi_J.
\end{equation}
This can then be generalized to any other loop configuration \cite{ricca1998applications,arnold2008topological}. Similar concepts (with different formulations) have appeared all the way from Gauss (1833) to topological field theory pioneered by Witten \cite{1989CMaPh.121..351W}.

In the standard fluid theory of vortices and in MHD the vortex linkage is conserved, i.e. $\dot H=0$. An early relativistic proof was given by Carter in~\cite{Carter:1979}, thereby foreshadowing Chern-Simons theory. Carter introduced a conserved vorticity four-vector current:
\begin{align}
    \mathcal{H}^0 &:= A_i B^i,\nonumber\\
    \mathcal{H}^i &:= \phi B^i + \mathbf{e}^{ijk} E_j A_k,
\end{align}
(with $\partial_\mu{\cal H}^\mu=0$ in the relevant cases). Up to a sign (depending on conventions), this current is precisely our ${\cal T}^\mu_{\Lambda}=- {\cal H}^\mu$, i.e. the dual of the Chern-Simons three-form, as stressed just after Eq.~(\ref{chernsimonscurrent}). A simple calculation reproduces Carter's argument in our language:
\begin{equation}
    \partial_{\mu} \mathcal{T}^{\mu}_{\Lambda} =
    \frac{1}{2}\mathcal{F}^{\alpha \beta} F_{\alpha \beta}=2 E_i \mathcal{B}^i.
\end{equation}
In the ideal cases considered in the MHD and fluids literature, the RHS vanishes \cite{moffatt_1969,berger1984topological}. For example, Ohm's law in the absence of resistivity $\rho$, implies $E_i + \mathbf{e}_{ijk} u^j B^k = 0$, and so $E_i B^i=0$.

\begin{figure}[h!]
    \centering

    \includegraphics[width=7cm]{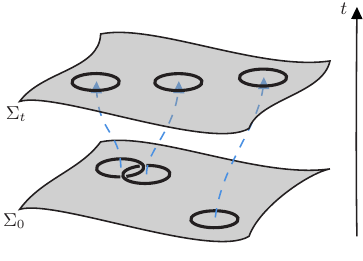}

    \caption{Spacetime volume forms between two hypersurfaces at different times, driven by changing magnetic flux tube configurations.}
    \label{fig:vortexlines}
\end{figure}

In contrast with the MHD setting, in our case conservation of linking number is undesirable. 
 Indeed, the passage of time is precisely to be interpreted as the creation (or annihilation) of links between vortices. This is illustrated in Fig.~\ref{fig:vortexlines}.
For unimodular time, we have
\begin{equation}
     \partial_{\mu} \mathcal{T}^{\mu}=2 E_i \mathcal{B}^i = \sqrt{-g}, 
\end{equation}
so that
\begin{align}\label{timederivativehelicity}
    \dot T = -\dot H = 2 \int_{\Sigma(t)} d^3x \ E_i \mathcal{B}^i = \int_{\Sigma(t)} d^3x \ N \sqrt{h}.
\end{align}
Similar expressions apply to other time measures, such as Ricci time. Obviously there is helicity conservation if time stops (e.g. for degenerate metrics or with vanishing Ricci for the two clocks we have examined), but not if the clocks are ticking.


It would be tempting to interpret this non-conservation by appealing to the MHD analogy: to see the passage of time as the result of resistivity, $\rho>0$. However,  this does not make sense. As pointed out at the end of the previous section, the analogy with EM does not extend to non-homogeneous Maxwell's equations and electric currents. Thus, the concept of resistivity has no role to play in this theory, at least in its present form. 

Another theoretical bridge that does not necessarily apply is with topological field theory~\cite{1989CMaPh.121..351W}. There, one is interested in non-simply connected leaves  $\Sigma_t$, so that large gauge transformations can be used to work out the Chern-Simons functional (i.e. the helicity with an appropriate normalization). The fact that this produces results modulo $2\pi$ reduces the gauge invariant configurations to $[0,2\pi)$. This point has been recognized in the MHD community~\cite{berger1984topological}, but often such situations are dismissed as artificial in that context, with the topology of space and boundary conditions chosen so that the helicity is directly gauge-invariant (the matter warrants a single line in~\cite{moffatt1992helicity}). In syntony with this work we will assume spatially simply connected cosmological models, except for a brief discussion of exceptions in the last Section.


\section{Topological discretization of time}\label{sec4}

We thus arrive at a picture where time is a helicity, which in turn implies a configuration of $\mathcal{B}^i$-vortices so that the value of time is related to {\it both} the linkage of these vortices {\it as well as} their fluxes. As can be seen from \eqref{magnetichelicity}, a priori a large number of configurations of vortices with different fluxes $\Phi_i$ and linking numbers can correspond to the same helicity. Furthermore, in spite of the discrete nature of the linking number, there is no automatic discretization of the helicity, and so of time, because the flux can be continuous. 

This discretization only arises if the flux in each vortex is assumed to be a Universal constant, $\Phi_0$, its fixed amount determining the quantum of time (which could be the Planck time). Let us assume a compact set of $\Sigma_t$. In each of them there is a discrete number of vortices and these have a discrete linking number. If the flux in each vortex is fixed, time is then the linking number times this flux.

With this additional postulate, time is identified with the linking number of vortices of magnetic fields $\mathcal{B}^i$: the passage of time becomes  the creation (or annihilation, depending on a sign convention) of new links. As a result, time is discretized, that is, it can only vary by an integer number of times of the fundamental flux $\Phi_0$:
\begin{equation}\label{discretetime}
    T_{\Lambda}(N) = - N \Phi_0^2, \ \ \ \ N \in \mathbb{Z}.
\end{equation}
In turn, one has a discretization of spacetime volume on-shell:
\begin{equation}\label{discretevolume}
    V_4(\Sigma_t) = - N(\Sigma_t) \Phi_0^2 V_3,
\end{equation}
for the case of unimodular time. This is true if the spatial leaves have finite volume~\footnote{The argument can be generalized to non-compact $\Sigma_t$ with a limiting procedure whilst keeping time intensive (see discussion around \eqref{intensivetimes}). In that case we should assume that the spatial density of vortices is constant.}.

Notice that within the sequester model \cite{JoaoPaper,JoaoLetter,Magueijo:2023poa,pad,pad1}, with basis given in (\ref{alphadef}), one can identify the observed value of the cosmological constant with the ratio of integers times the square of the fundamental fluxes associated with each constant of Nature:
\begin{equation}\label{observedcosmologicalconstant}
    \Lambda_{\text{obs}} \sim - \frac{1}{4} \frac{N_{\Lambda}(\Sigma_t)}{N_R(\Sigma_t)} \bigg(\frac{\Phi_{0,\Lambda}}{\Phi_{0,R}}\bigg)^2.
\end{equation}
The importance of this result comes from the quantum nature of the fluxes and might explain the smallness of the observed value of the cosmological constant by taking $\Phi_{0,\Lambda}$ to be of the order of the Planck scale, i.e. $\Phi_{0,\Lambda}^2 \approx 3.16 \times 10^{-33} \ Wb^2$.

\section{Variable constants and degenerate metrics}\label{sec5}


A more subtle point appears when solving (\ref{constantlambda}). It was shown in \cite{Bansal:2021bis} that its most general solution, assuming $\partial_{\mu} \Lambda \neq 0$, is
\begin{equation}\label{vectorpotentialdecomp}
    A_{\mu} = \partial_{\mu} a + b \partial_{\mu} \Lambda, \ \ \ \ a,b \in \mathcal{C}^{\infty}(\mathcal{M}).
\end{equation}
An important consequence of this solution involves having a good choice of global coordinate $\Lambda$: the level surfaces of constant $\Lambda$ provide a globally nondegenerate foliation of $\mathbb{R}^4$, with each slice having topology $\mathbb{R}^3$. However, we can also have solutions to (\ref{constantlambda}) using the invertibility of the dual-field strength, which makes $\Lambda = \text{const.}$ trivially true. Thus, there is a class of solutions with constant $\Lambda$, but is this the most general solution?

In fact, (\ref{constantlambda}) can be solved. First, the determinant of the field strength satisfies the standard identities:
\begin{equation}
    \text{det}({\cal F}_{\mu \nu})= \text{det}(F_{\mu \nu}) = (E_i \mathcal{B}^i)^2 \equiv \frac{1}{16} (\mathcal{F}^{\mu \nu} F_{\mu \nu})^2.
\end{equation}
Hence (\ref{unimodularcondition}) implies:
\begin{equation}\label{relatedets}
    \sqrt{-g} =  2 \sqrt{\text{det}(F_{\mu \nu})}
    =2 \sqrt{\text{det}({\cal F}_{\mu \nu})}.
\end{equation}
Wherever the metric is non-degenerate ($g\neq 0$), we therefore have $\text{det}({\cal F}_{\mu \nu})\neq 0$, so that ${\cal F}_{\mu \nu}$ is invertible, and Eq.~(\ref{constantlambda}) implies $\partial_\mu\Lambda=0$. Hence we need regions with degenerate metric (and consequent degenerate 
$\mathcal{F}^{\mu}_{\ \nu}$)
for $\Lambda$ to be allowed to vary. This implies that $E_i \mathcal{B}^i=0$, i.e. that the electric and magnetic fields are orthogonal, but closer inspection shows that the condition is more restrictive. What follows lifts results to be found in \cite{hacyan2011algebraic}, where the eigenvalue problem
\begin{equation}\label{eigenvaluedualF}
    \mathcal{F}^{\mu}_{\ \nu} k^{\nu} = \lambda k^{\mu},
\end{equation}
was examined. The fact that $\mathcal{F}^{\mu}_{\ \nu}$ is anti-symmetric implies that the 
eigenvectors  are null $k^{\mu} k_{\mu} = 0$. The characteristic polynomial leads to:
\begin{equation}\label{characteristicpoly}
    \lambda^4 - (\mathcal{B}_i \mathcal{B}^i - E_i E^i) \lambda^2 - (E_i \mathcal{B}^i)^2 = 0.
\end{equation}
There are four solutions to this equation: $\lambda = \pm i \sqrt{E_i E^i}$ or $\lambda = \pm \sqrt{\mathcal{B}_i \mathcal{B}^i}$. Hence, for there to be a null eigenvalue we need for either the magnetic or the electric field to vanish (or both). In the first case, there would be no vortices at all in our space, in contradiction with the postulate of the last Section. In the second, there are vortices but no electric field is present over a leaf (or set of leaves). This is equivalent to the MHD situation, even though it is not justified by zero resistivity (a concept that does not make sense in this context, as explained before). The Bianchi identities imply that $\dot {\mathcal{B}^i}=0$, so the vortices are actually frozen-in. As a corollary, the linking number is trivially conserved and time does not flow. Concomitantly, $\Lambda$ is allowed to change.

The above applies to $\boldsymbol{\alpha}=\alpha_1=\rho_\Lambda$ (using the basis (\ref{alphadef})), but similar conclusions can be drawn for other target constants and their respective times (and types of vortices). For instance, taking the $\alpha_2$ variable, we have:
\begin{equation}\label{Ricciflat}
    -\frac{1}{2} R \sqrt{-g} = \pm 2 \sqrt{\text{det}\big(F_{\mu\nu}^{\alpha_2}\big)}. 
\end{equation}
The Planck mass can now change in Ricci flat leaves, or in leaves where the integrated Ricci scalar vanishes  (for example in a pure radiation dominated Universe, under strict conformal invariance). Again conservation of linking number (as well as zero electric field, and no time variation in the magnetic fields) is restored for the corresponding vortices in this regime.

Finally, how can the parameters $\boldsymbol{\alpha}$ vary in such cases? We have to reconcile the Bianchi identities with (\ref{constantlambda}): 
\begin{align}
      &\mathcal{B}^i \partial_i \boldsymbol{\alpha} = 0 \nonumber\\
      &\mathcal{B}^i \dot{\boldsymbol{\alpha}} + \mathbf{e}^{ijk} (\partial_j \boldsymbol{\alpha}) E_k=0,
\end{align}
(which would suggest looking at $\boldsymbol{\alpha}$ as an electric permittivity and magnetic permeability, except that we are looking for constraints on $\boldsymbol{\alpha}$). As we know we must choose which field to set to zero to allow variability. If we set $E_i=0$ in accordance to our postulate, we get that the variation must be purely spatial, and orthogonal to the magnetic flux tubes.



\section{Non-abelian extension}\label{sec7}


We now aim to extend the reduced theory by replacing the divergence in (\ref{fullaction}) with a non-abelian Chern-Pontryagin term. This results in the following identification, where the action exhibits a symmetry group $G$:
\begin{equation}\label{nonabelianaction}
    \int d^4x \ \Lambda \partial_{\mu} \mathcal{T}^{\mu}_{\Lambda} = \int d^4x \ \frac{\Lambda}{2} \mathcal{F}^{a \mu \nu} F^a_{\mu \nu},
\end{equation}
As with the abelian case in (\ref{chernsimonscurrent}), we relate:
\begin{equation}\label{nonabeliancurrent}
    \mathcal{T}^{\alpha} = e^{\alpha \beta \mu \nu} \bigg(A^a_{\beta} \partial_{\mu} A^a_{\nu} + \frac{1}{3} f^{abc} A^a_{\beta} A^b_{\mu} A^c_{\nu} \bigg),
\end{equation}
which is the one-form dual to the non-abelian Chern-Simons three-form, i.e. the Chern-Simons current.

With (\ref{nonabeliancurrent}) implying 
\begin{equation}
    \mathcal{T}^{0} = - \mathbf{e}^{ijk} \bigg(A^a_{i} \partial_{j} A^a_{k} + \frac{1}{3} f^{abc} A^a_{i} A^b_{j} A^c_{k} \bigg),
\end{equation}
we obtain the unimodular time from (\ref{unimodulartime}):
\begin{align}\label{nonabelianunimodulartime}
    T_{\Lambda}(t) &= \int_{\Sigma_t} d^3x \ \mathcal{T}^0\nonumber\\
    &= - \int_{\Sigma_t} d^3x \ \mathbf{e}^{ijk} \bigg( A_i^a \partial_{j} A_{k}^a + \frac{1}{3} f^{abc} A_{i}^a A_{j}^b A_{k}^c \bigg).
\end{align}
This is nothing but the Chern-Simons number defined over a three-volume $V$ (up to a sign):
\begin{equation}\label{Chernsimonsnumber}
    CS_3(A_i) := \int_V d^3x \ \mathbf{e}^{ijk} \bigg( A_i^a \partial_{j} A_{k}^a + \frac{1}{3} f^{abc} A_{i}^a A_{j}^b A_{k}^c \bigg).
\end{equation}
This coincides with a \textit{non-abelian helicity} as follows. By introducing a ``densitized chromomagnetic field" $\mathcal{B}^{ai}$:
\begin{align}\label{nonabelianmagneticfield}
    \mathcal{B}^{ai} &:= \mathbf{e}^{ijk} \bigg(\partial_{j} A_{k}^a + \frac{1}{3}f^{abc} A_{j}^b A_{k}^c \bigg)\nonumber\\
    &\equiv \mathcal{B}^{ai}_{\text{A}} + \frac{1}{3} \mathcal{G}^{ai},
\end{align}
where $\mathcal{B}^{ai}_{\text{A}}:= \mathbf{e}^{ijk} \partial_{j} A_{k}^a$ represents $a$-independent abelian magnetic fields, and $\mathcal{G}^{ai}:= f^{abc} \boldsymbol{e}^{ijk} A_{j}^b A_{k}^c$ expressing the interaction among the different colors of the gauge group, one can rewrite (\ref{Chernsimonsnumber}) as
\begin{equation}
    CS_3(A_i) = \int_{V} d^3x \ A_i^a \mathcal{B}^{ai} \equiv H_{\text{CS}}
\end{equation}
with the non-abelian extension of helicity:
\begin{equation}\label{nonabelianhelicity}
    H_{\text{CS}} := \sum_{a=1}^{\text{dim}(G)} H^a + \frac{1}{3} \int_V d^3x \, A_i^a \mathcal{G}^{ai}.
\end{equation}

\begin{figure}[h]
    \centering
    \includegraphics[width=7cm]{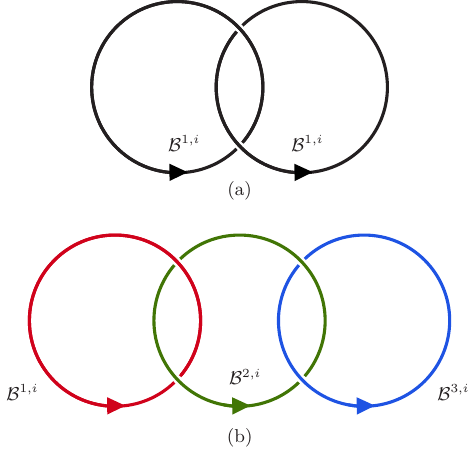}
    
    \caption{Non-abelian vortices in $SU(2)$: In (a), a Hopf link formed from the same color vortices, with abelian helicity measuring their linkage. In (b), a complex link of different color vortices interlinks, requiring a generalized cross-helicity to calculate a third-order linking number.}
    \label{fig:nonabelianlinks}
\end{figure}

What does this extended helicity correspond to in terms of unimodular time? As for the abelian case, we have $\mathcal{B}^{ai}$-vortex lines, with the additional complication of different gauge-colored vortices interacting among themselves. For example, in the context of the gauge symmetry $SU(2)$, we anticipate the existence of three distinct types of vortices, which encompass both free and interacting chromomagnetic field vortices (refer to Figure \ref{fig:nonabelianlinks}).
Specifically, in (\ref{nonabelianmagneticfield}), the interaction term $\mathcal{G}^{1i} := 2 \mathbf{e}^{ijk} A_{j}^2 A_{k}^3$ (along with its cyclic permutations in the color indices) gives rise to a third-order topological invariant, akin to cross-helicity, explored in previous works \cite{hornig2002towards, mayer2003topological}. This can be expressed as:
\begin{equation}\label{generalizedcrosshelicity}
H^{(3)}(\mathcal{B}^{1i}_{\text{A}},\mathcal{B}^{2i}_{\text{A}},\mathcal{B}^{3i}_{\text{A}}) := \int_V d^3x \  A_i^1 \mathcal{G}^{1i} \equiv H^{123}.
\end{equation}
This invariant satisfies $H^{123} = H^{312} = H^{231}$. Similar third-order link integrals, describing vortex linkage, have been studied in previous research \cite{berger1990third, evans1992hierarchy, bodecker2004link}.

\section{Conclusions and outlook}

In this paper, we studied the Henneaux-Teitelboim version of unimodular gravity and its axion reduction. This amounted to replacing the total divergence (the 4-form) that appears in the unimodular action with the $\theta$-term found in axion models \cite{vikman-axion}. This axionic theory has fewer gauge degrees of freedom than HT gravity:  we go from three to one. 
We stress that this abelian reduced theory does not correspond to standard electromagnetism, because there is no non-topological term in the action, and no non-homogeneous Maxwell equations present.

The core result of this paper hinges on the identification of unimodular time with the helicity in our reduced theory, mirroring the abelian Chern-Simons term.
Using standard results, time is thus identified with the linking number of the magnetic field vortices multiplied by their fluxes.
In principle, numerous vortex configurations with different fluxes and linking numbers can yield the same helicity. We postulated that all vortices carry the same Planck-scale magnetic flux. Consequently, time is identified with the linking number, introducing a discrete nature, whereby it can vary only in integer multiples of the fundamental flux. 

We close this paper by discussing potential future directions. 
Foremost the blatant connection with the sequester model~\cite{pad, pad1,JoaoPaper, JoaoLetter, Magueijo:2023poa} should be followed up, since it relates our considerations with the observed cosmological constant's value. In this context, cross-helicity, a consequence of the $U(1) \times U(1)$ symmetry in the action, could have a role to play, relating the finite 4-volume and Ricci fluxes, and so providing an explanation for the value of the stabilized cosmological constant. More generally, 
we have assumed in this paper that the leaves $\Sigma_t$ are simply connected, but what if they are not? The gauge-invariant helicities would then be naturally bounded, providing justification for one of the assumptions of sequestration. Could the topology of space fix the topology of time? And if time is cyclic, what would be the implication for thermal states in the early Universe?

Another avenue for exploration involves incorporating gravitational effects into the electric and magnetic fields, either through gravitational deformations of BF theory or alternative topological approaches to gravity. Relation with topological field theory, and how the usual 2+1 set up translates into 3+1, should also be explored. Additionally, the reduction aspect of the theory should be thoroughly examined using the Stuckelberg procedure in future work. Notably, the emergence and physical interpretation of helicity depends significantly on the underlying base theory. Thus, it is essential to analyze these aspects in the context of other base actions, including Brans-Dicke theory, modified gravity theories, higher-derivative theories, massive gravity, and even supergravity. Lastly, we should also consider the potential implications of this model on gravitational solutions, such as black holes and cosmology.

\section{Acknowledgments}
We thank Amihay Hanany, Raymond Isichei, Karapet Mkrtchyan, and Daniel Waldram for helpful discussions related to this paper. The work of JM was partly supported by the STFC Consolidated Grants ST/T000791/1 and ST/X00575/1.



\bibliographystyle{apsrev4-2}
\bibliography{references}

\end{document}